\documentstyle[sprocl]{article}

\def\be{\begin{equation}} 
\def\ee{\end{equation}}
\def\bea{\begin{eqnarray}} 
\def\eea{\end{eqnarray}} 

\begin{document}

\begin{flushright} 
\begin{tabular}{l} 
VPI-IPPAP-98-4 \\ 
hep-th/9807084 
\end{tabular} 
\end{flushright}

\vspace{8mm}

\begin{center}

{\Large \bf Black hole in de Sitter space}\footnote{Talk presented 
  at the Sixth International Symposium on Particles, Strings and Cosmology ({\bf
  PASCOS-98}), Northeastern University, March 22-29, 1998.}

\vspace{8mm}

{\large Feng-Li Lin}\footnote{ The author would like to thank Dr. Chopin 
Soo for useful comments on the manuscript.}
\footnote{ email address:  linfl@vt.edu}

\vspace{2mm} 
Department of Physics, and Institute for Particle Physics and
Astrophysics, Virginia Polytechnic Institute and State University, Blacksburg,
VA 24061-0435, U.S.A.  \\



\vspace{15mm}

{\bf Abstract}

\end{center}

  If cosmological constant is positive, a black hole is naturally described by
the Schwarzschild-de Sitter solution with two horizons.  We use the global
method to extract the topological information and the selection rule for the
Gibbons-Hawking temperature for the thermal vacua.  These are related to the
Euler number of the Euclidean section whose topology is more complicated than
expected.  We also point out the failure of the usual local method of conical
singularity approach in dealing with multi-horizon scenarios.

\vspace{45mm}

\begin{flushleft} 
\begin{tabular}{l} 
VPI-IPPAP-98-4 \\ July, 1998 
\end{tabular}
\end{flushleft}

\vfill

\pagestyle{empty}

\pagebreak

\pagestyle{plain} \pagenumbering{arabic}

\title{Black hole in de Sitter space }

\author{Feng-Li Lin}

\address{ Department of Physics, and Institute for Particle Physics and
Astrophysics, Virginia Polytechnic Institute and State University, Blacksburg,
VA 24061-0435, U.S.A.  \\E-mail:  linfl@vt.edu}


\maketitle 
\abstracts{If cosmological constant is positive, a black hole is
naturally described by the Schwarzschild-de Sitter solution with two horizons.
We use the global method to extract the topological information and the
selection rule for the Gibbons-Hawking temperature for the thermal vacua. These
are related to the Euler number of the Euclidean section whose topology is more
complicated than expected.  We also point out the failure of the usual local
method of conical singularity approach in dealing with multi-horizon scenarios.}

 The evidence of the positive cosmological constant lead us to wonder how the
usual black hole thermodynamics will be modified in this non-asymptotically flat
de Sitter background.  In this regard, we have to account for both the
cosmological as well as the black hole horizons.  It seems that no global
thermal temperature could be defined unless these two horizons are coincident,
however, then the Euclidean section in between the horizons shrink to zero, and
not physically interesting.  We may then ask if it is possible to define a 
global and unique Gibbons-Hawking temperature\cite{GH} and the associated 
thermal vacuum in the generic static multi-horizon scenario.  The answer is 
surprisingly in the affirmative\cite{LS}.  

  For the Schwarzschild black hole, the Hawking temperature is defined by
requiring the removal of the conical singularity on the horizon.  So unlike the
quantum theory on flat $R^3\times S^1$ Euclidean background in which the thermal
temperature is just a parameter and could be any values, in the curved
background with horizon, the thermal temperature and thus the vacuum are
prescribed by the geometry and not arbitrary.  The conical method\cite{FS} can
not be applied to the multi-horizon spacetimes, such as the Schwarzschild-de
Sitter spacetime, because one can not find a temperature to remove the conical
singularities on both horizons simultaneously.  If so, the one-loop quantum
theory will blow up on the conical singularities at horizons, and be not
well-defined.  The more deep reason is that the local geometry information from
conical singularity around each horizon is not enough to determine the global
feature such as temperature in the multi-horizon scenario.  This reminds us the
similar situation about the singular string of the Dirac monopole, and implies a
global method with the patching conditions is needed in dealing with the
nontrivial topology with the multi-horizon scenarios.

  It turns out the Kruskal extension is the proper tool to analytically continue
beyond one horizon, and at the same time carry the information to another 
horizon for precise matching. To be more specific, let us first consider the 
pure Schwarzschild black hole. The metric is
\begin{equation}
ds^2= -(1-\frac{2m}{r})dt^2 + \frac{dr^2}{1-\frac{2m}{r}} + r^2 d\Omega^2
\end{equation}
We then define the tortoise coordinate and perform the conformal 
transformation to remove the coordinate singularity at the horizon, after all
are done the metric becomes
\begin{equation}
ds^2 ={1\over {k^2r}}e^{-2kr}({dr'}^2 -{dt'}^2)  + r^2 d\Omega^2 \;.
\end{equation}
Here the surface gravity $k=1/4m$. Now the metric is regular at $r=2m$ and the 
old and new coordinates are related by
\begin{eqnarray}
{r'}^2 - {t'}^2 &=&  e^{2kr}(r - 2m)  \\
t'&=& e^{k r^*}\sinh(-ik\tau)  
\end{eqnarray}
After Euclideanization by $t=i\tau$, we could see that Eq.(3) define the
Euclidean section to be $r>2m$, and Eq.(4) forces $\tau$ to have minimal 
period equal to the inverse Hawking temperature. So we see the global method 
yield the same Hawking temperature as the one by local conical method, and also
give the global information about the region of the Euclidean section.

  Next we move on to the Schwarzschild-de Sitter case, the metric is
\begin{equation}
ds^2 = -h(r)dt^2 + h^{-1}(r)dr^2 + r^2 d\Omega^2 \; ,
\end{equation}
with
\begin{equation}
h(r) = {\lambda \over {3r}}(r_+ - r) (r- r_-)(r+ r_+ + r_-).
\end{equation}
It is a solution of Einstein's equations with cosmological constant 
$\lambda$ if
\begin{equation}
3/\lambda = r^2_+ + r_+r_- + r^2_- ,\qquad 6m/\lambda = r_+r_-(r_+ + r_-).
\end{equation}
The metric is a priori defined for the region between the horizons but there can
be a Kruskal extension.\cite{TT} Because there are two horizons, we need the
separate coordinate patches around each respective horizon to complete the
Kruskal extension.  After continuation, the metric for each patch becomes
\begin{equation}
ds^2=h_{\pm}(-dt'^2_{\pm} + dr'^2_{\pm}) + r^2 d\Omega^2 \; .
\end{equation}
Here $h_+$($h_-$) is regularly defined only for $r>r_-$($r<r+$). This defines 
the Euclidean section to be $r_-<r<r_+$. And the new and old time coordinates
on each patch are related by
\begin{equation}
t'_{\pm} = e^{\mp k_{\pm} r^*} sinh(k_{\pm} t)
\end{equation}
Here the tortoise coordinate $r^*= \int h^{-1}(r) dr$, and the surface gravity
on each horizon $k_{\pm}=1/2r_{\pm}$.  After Euclideanization, Eq.(9) will force
$\tau=-it$ to have a minimal period as the lowest common multiple of
$2\pi/k_{\pm}$ as long as the ratio $\alpha \equiv \frac{k_+}{k_-} = \frac{n_+}{
n_-}$ is a rational number($n_{\pm}$ are prime integers).  From the minimal
period $\beta = \frac{2\pi n_{\pm}}{k_{\pm}}$, we then define a global
Gibbons-Hawking temperature to be
\begin{equation}
T_{GH}  \equiv \frac{1}{\beta} = \frac{k_+ + k_-}{2\pi(n_++n_-)}
\end{equation}
We can also calculate the transition function between the two patches to check 
that it is well-defined. As $\tau$ is shifted by $\beta$, the 
transition function is changed by a phase $e^{i 2\pi (n_+ +n_-)}=1$, and thus
unique. This shows that our patching is consistent and $n_+ + n_-$ is the 
winding number of the patching. 
Note that $T_{GH}$ is defined only for prime integers of $n_{\pm}$, so it is
in discrete values in order to have a thermal vacuum.  This selection rule of
the thermal vacuum is further exemplified by calculating the Euler number of the
Euclidean section
\begin{equation}
\chi[M]= {1 \over 32\pi^2}\int_{M} d^4x {\sqrt g}
(R^2-4R_{\mu \nu}^2+R_{\mu \nu \alpha \beta}^2),
\end{equation}
which turns out to be equal to $2(n_+ + n_-)$, and shows that the topology of 
the Schwarzschild-de Sitter spacetime is more complicated than one would expect. 
The thermal physics of the Schwarzschild-de Sitter space is closely
correlated with the topology of the spacetime through the formula
\begin{equation}
T_{GH} = \frac{k_+ + k_-}{\pi \chi}
\end{equation}
 This is the generalization of $T=\frac{k}{2\pi}$ for the pure black hole or de
Sitter spacetimes in our proposal of incorporating the horizons via the Kruskal
extension, and is generalizable to even more complicated multi-horizon
scenarios.

\end{document}